\documentclass[11pt,twoside]{article}

\usepackage{asp2006}
\usepackage{epsf}
\usepackage{psfig}
\usepackage{lscape}

\begin{document}
\title{Towards simulating the photometry, chemistry, mass loss and 
pulsational properties of AGB star populations in resolved galaxies} 
\author{L\'eo Girardi\altaffilmark{1} \& Paola Marigo\altaffilmark{2}}
\altaffiltext{1}{Osservatorio Astronomico di Padova, 
Padova, Italy}
\altaffiltext{2}{Dipartimento di Astronomia, Universit\`a di Padova, 
Padova, Italy}

\begin{abstract} 
Extended and updated grids of TP-AGB tracks have been implemented in
the TRILEGAL population synthesis code, which generates mock stellar
catalogues for a galaxy given its mass, distance, star formation
history and age-metallicity relation, including also the Milky Way
foreground population. Among the stellar parameters that are
simulated, we now include the surface chemistry, mass-loss rates,
pulsation modes and periods of LPVs.  This allows us to perform a
series of consistency checks between AGB model predictions and
observations, that we are just starting to explore. We present a few
examples of model--data comparisons, mostly regarding the
near-infrared and variability data for AGB stars in the Magellanic
Clouds.
\end{abstract}


\section{Introduction} 
\label{sec_intro}

Synthetic evolutionary models play an important role in the study of
TP-AGB stars. The synthetic approach allows the computation of
evolutionary grids covering wide age and metallicity intervals, and
the easy testing of different input prescriptions (like the mass loss
rates and dredge-up efficiency) so as to reproduce basic observational
constraints.  Due to these characteristics, they are suitable for
population synthesis of galaxies, i.e. to the modelling of their
resolved stars, and of their integrated light.

Since the work by Groenewegen \& de Jong (1993), the carbon star
luminosity functions (CSLFs) in the Magellanic Clouds (MCs) have been
used to calibrate the onset and efficiency of third dredge-up events
in synthetic TP-AGB models.  Nowadays, many more observables
potentially useful to the calibration process of synthetic models can
be individuated. Suffice it to recall that for a few regions of the
MCs, we dispose of optical plus near-IR photometry of all stars above
the RGB-tip (OGLE, DENIS, 2MASS), spectroscopic C/M classification,
variability data (EROS, OGLE, MACHO), mid-IR photometry and spectra
(ISO, Spitzer), a good hint of their star formation history from deep
HST data, plus the [Fe/H] distribution of red giants from ground-based
spectroscopy.  The simulation of such observables starting from grids
of AGB tracks -- and taking into account biases, errors,
incompleteness, etc. -- should allow us to make many specific tests on
their processes of nuclear burning, mixing, mass-loss and pulsation,
and test correlations between quantities (e.g. $L$, $T_{\rm eff}$,
surface C/O ratio, mass loss rates, periods, etc.) that are expected
to be closely interrelated.


Therefore, we aim at a more complete testing and exploitation of
TP-AGB evolutionary tracks, by means of comparisons of mock catalogues
with real data for Local Group galaxies. To this aim, the TRILEGAL
code has been adapted to deal with sophisticated TP-AGB tracks
(Sect.~\ref{sec_tracks}) and to simulate more observables in addition
to the photometry -- including the spectral classification, mass loss
rates, and pulsational properties of AGB stars
(Sect.~\ref{sec_adapt}). A few preliminary applications are presented
in Sect.~\ref{sec_results}

\section{Grids of TP-AGB tracks and isochrones}
\label{sec_tracks}

The TRILEGAL code (Groenewegen et al.~2002; Girardi et al.~2005;
Vanhollebeke et al., this meeting) is a complex tool to produce mock
catalogues for both the Milky Way and external galaxies. It stands on
a series of routines that efficiently construct sample of stars by
interpolating within grids of evolutionary models, and then it
attributes a photometry to each star by consulting into tables of
bolometric corrections. Presently, there is a quite good calibration
for the Milky Way geometry in TRILEGAL, and external galaxies like the
MCs can be easily simulated for a fixed distance and foreground
reddening.  An interactive web interface to TRILEGAL is available at
{\tt http://trilegal.ster.kuleuven.be}.

\begin{figure}[!ht]
\plottwo{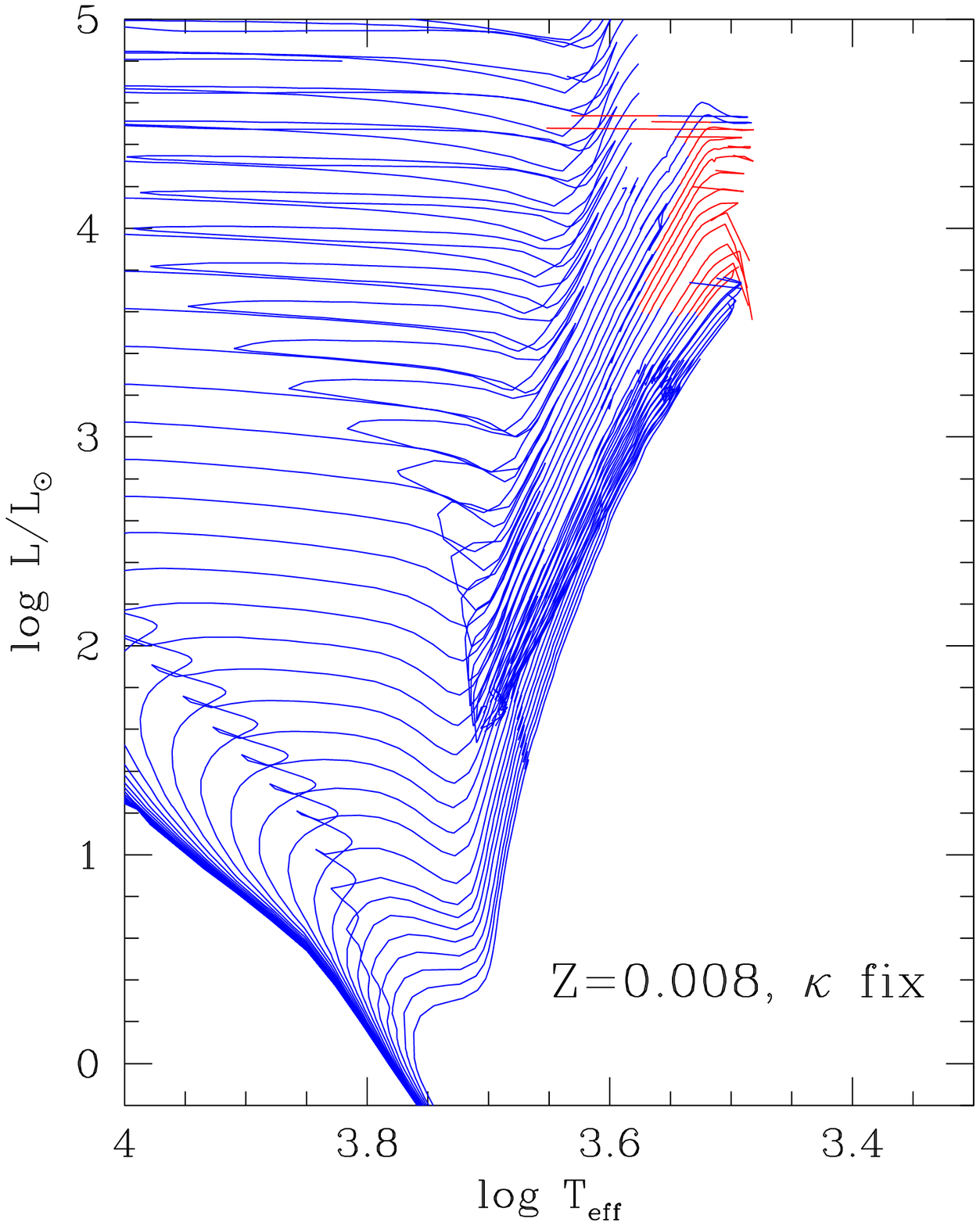}{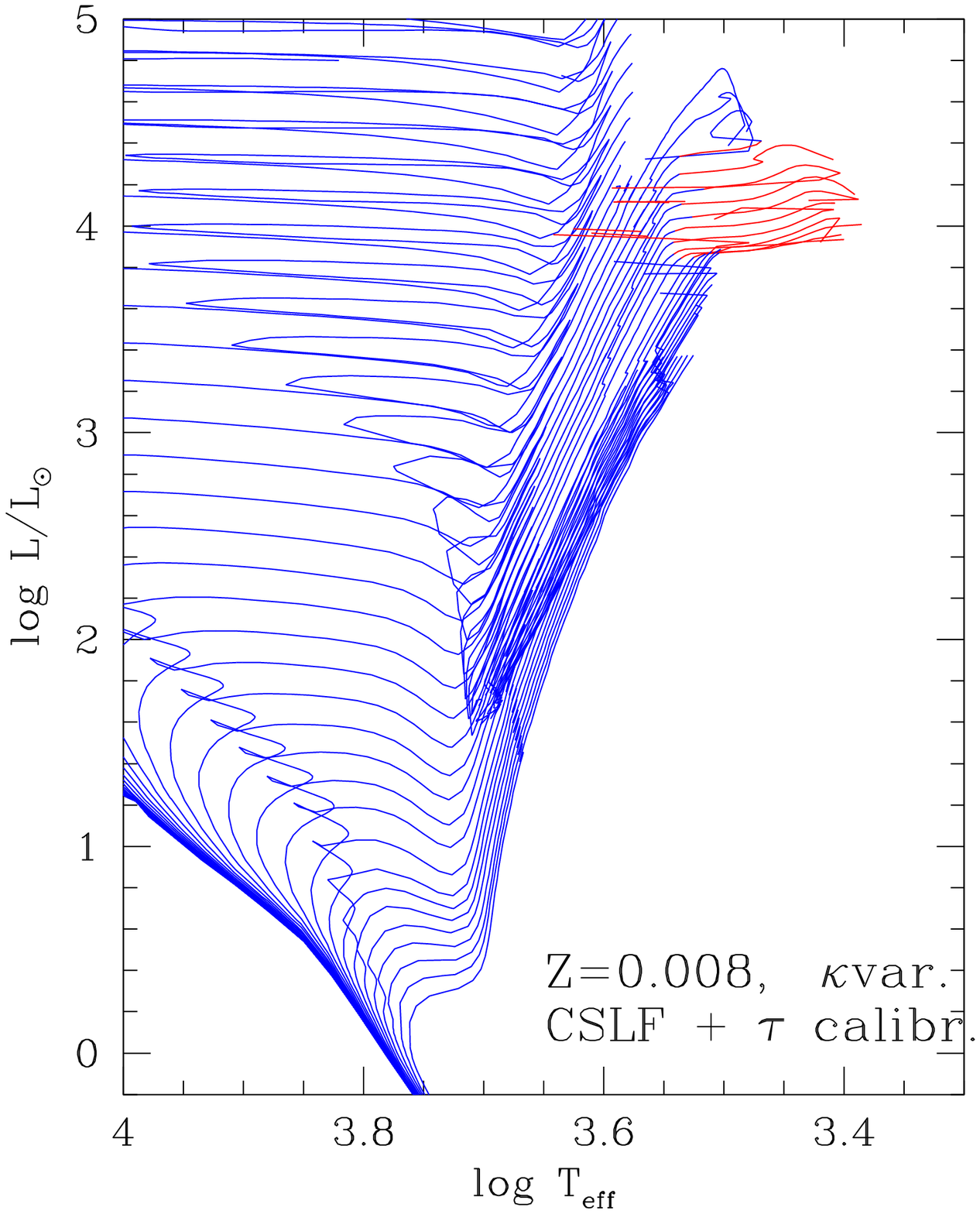}
\caption{Two of the isochrone sets already in use in TRILEGAL: the 
Marigo \& Girardi (2001; left) ones which use scaled-solar opacities
for the TP-AGB, and the latest Marigo et al.~(in prep.; right) using
variable opacities. 
}
\label{fig_isoc}
\end{figure}

TRILEGAL uses the low- and intermediate-mass stellar tracks by Girardi
et al.~(2000) up to the end of the early-AGB. These are complemented
with synthetic TP-AGB tracks from various sources:
(i) The same ones as described in Girardi \& Bertelli (1998), relying
on an extremely simplified formalism, without any dredge up and
hot-bottom burning; this naive approach was justified as initial
applications of TRILEGAL were not dealing with AGB stars.
(ii) Marigo et al.~(1999) tracks calibrated on the CSLFs in the MCs,
computed with -- as so far standard in the field of stellar evolution
-- the erroneous approximation of solar-scaled opacities; the
corresponding isochrones were published by Marigo \& Girardi (2001).
(iii) Tracks computed for studying the SFH in the Magellanic Clouds
(Cioni et al.~2006ab), now adopting consistent low-temperature
opacities for C-rich mixtures (the so-called $\kappa$-var case in
Marigo 2002).
(iv) Finally, a newly computed set of $\kappa$-var tracks (Marigo \&
Girardi, in prep.) calibrated not only on the CSLFs but also on the
lifetimes derived from the C- and M-star counts in MC star clusters
(Girardi \& Marigo 2006).

Some of these isochrone sets are illustrated in Fig.~\ref{fig_isoc}
(and are available in {\tt http://pleiadi.oapd.inaf.it}).  The
procedure for changing the set of TP-AGB tracks is straightforward so
that any future set of tracks can be immediately included and tested.
Notice that the isochrones of Fig.~\ref{fig_isoc} have been
constructed using the quiescent stages of H-burning along the
TP-AGB. This simplification is necessary to speed up the
interpolations inside grids of TP-AGB tracks. Individual tracks
instead have been computed with a detailed description of $L$ and
$T_{\rm eff}$ variations during thermal pulse cycles.

\section{Adapting TRILEGAL to the TP-AGB}
\label{sec_adapt}

A few features have been added to TRILEGAL, in order to face the
challenges posed by AGB populations:

First of all, in producing mock catalogues we re-introduce the $L$ and
$T_{\rm eff}$ variations driven by thermal pulses in AGB stars. This
is performed by generating a random phase of the pulse cycle for each
simulated star, then computing the $\Delta L$ variation with
Wagenhuber \& Groenewegen's (1998) formulas, and then converting it
into a $\Delta T_{\rm eff}$ by using the slopes in the HR diagram
previously derived from envelope integrations. This is performed
separately for C- and M-type stars, which present very different
$\Delta L/\Delta T_{\rm eff}$ slopes.

Second, we properly derive quantities such as the mass loss rates,
pulsation modes and periods for the interpolated stars. Our experience
is that this cannot be simply performed by interpolating between the
tracks, due to the recurrent switch of pulsation mode along the pulse
cycles. These quantities, instead, have to be re-derived from other
ones such as the stellar mass, radius, C/O ratio, etc., which are now
included into the tracks and isochrone tables.

Third, we need to properly deal with bolometric corrections for
C-stars, and for all AGB stars with dusty envelopes. This step is
presently under development. At the stage of this writing, we are
using empirical bolometric corrections and $T_{\rm eff}$-colour
relations for C-stars (cf. Marigo et al.~2003).


\section{A few results}
\label{sec_results}

\begin{figure}[!ht]
\plottwo{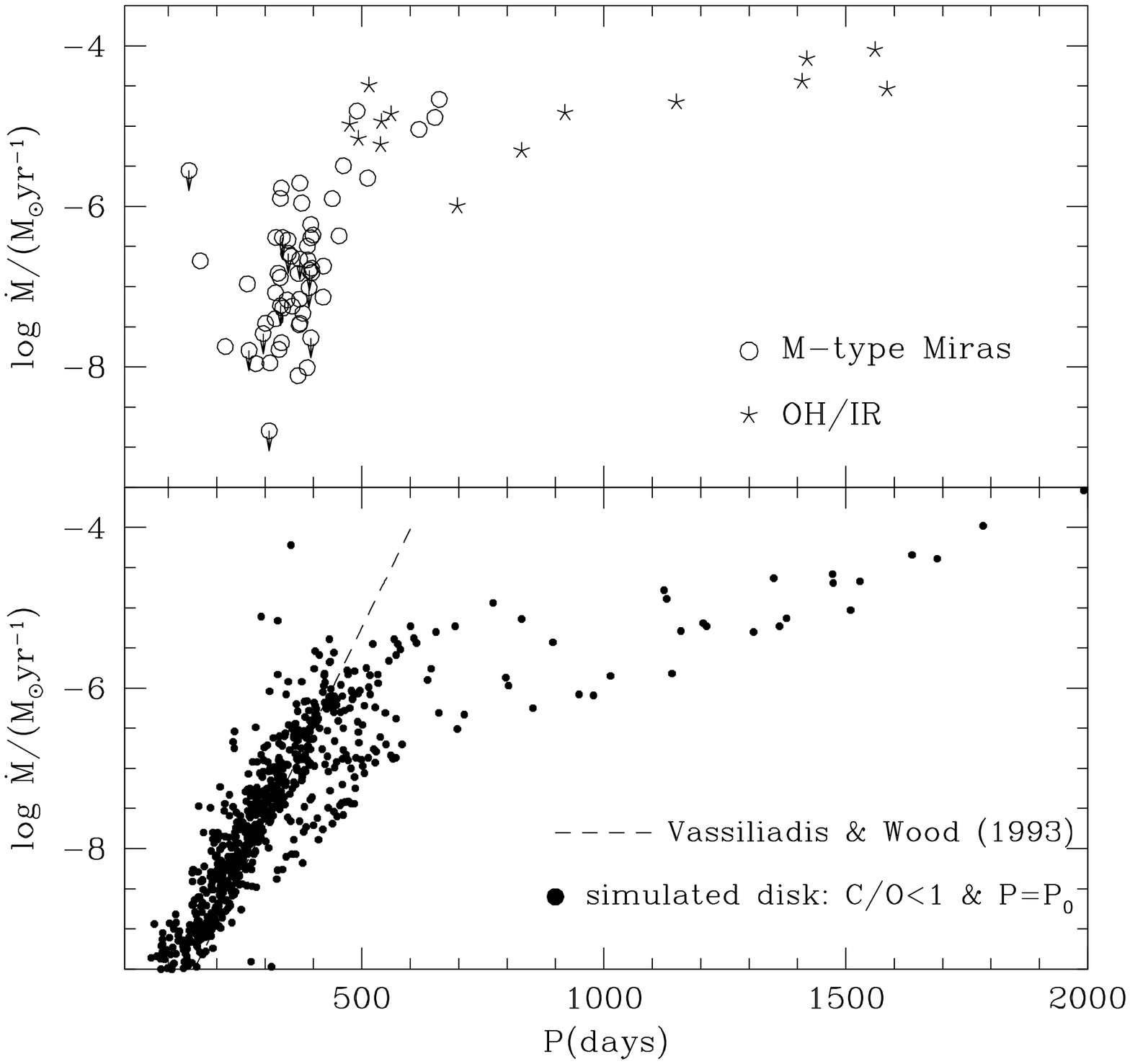}{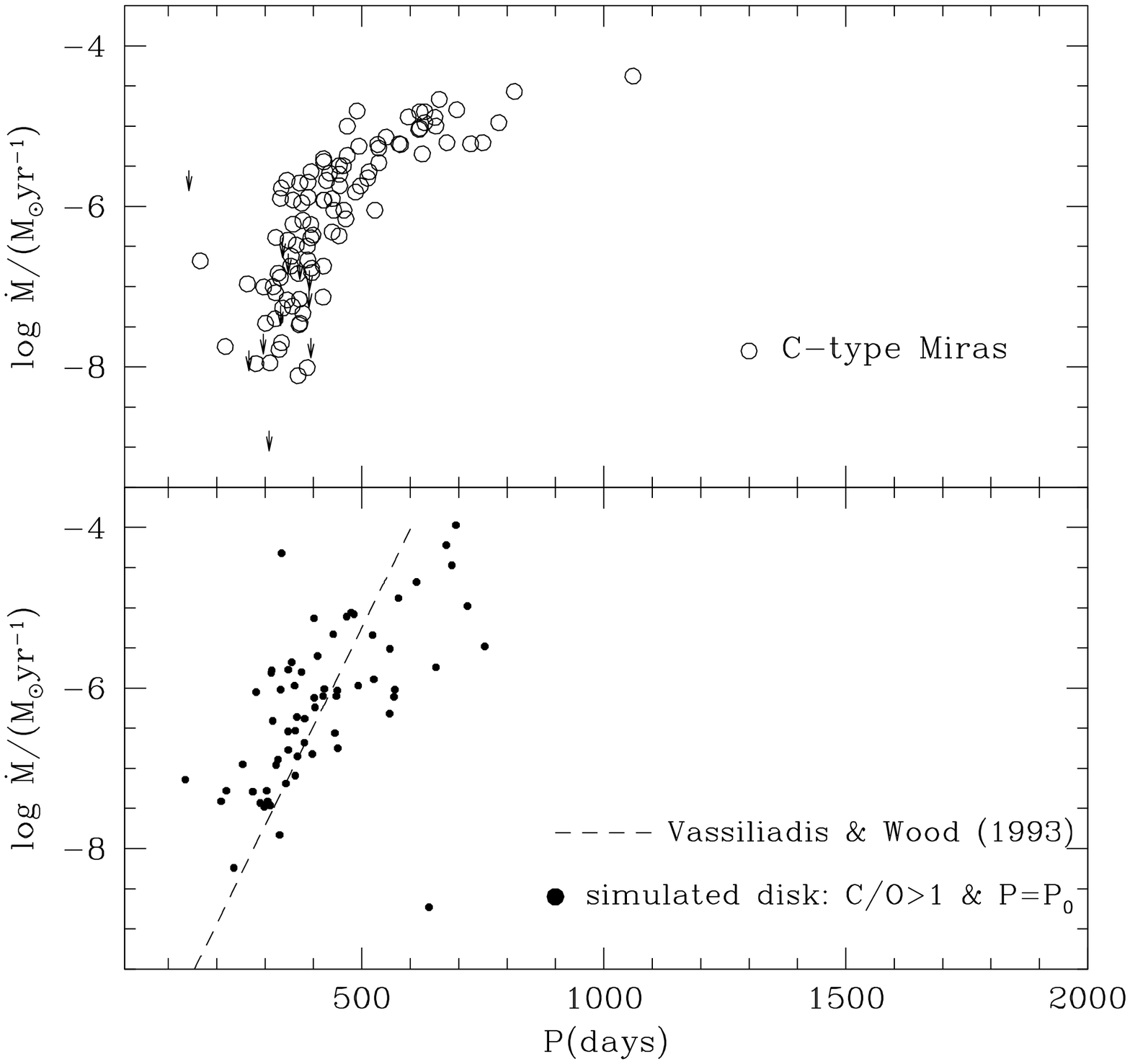}
\caption{Left panel: Observed (top; data from Le Bertre \& Winters 1998) and 
predicted (bottom) distribution of Galactic M-type Miras in the mass
loss vs. period diagram. Right panel: The same for C-type Miras (the
data is from Groenewegen et al.~1999). The dashed line in the bottom
panels represent the empirical relation by Vassiliadis \& Wood
(1993).}
\label{fig_mloss}
\end{figure}

\begin{figure}[!ht]
\plottwo{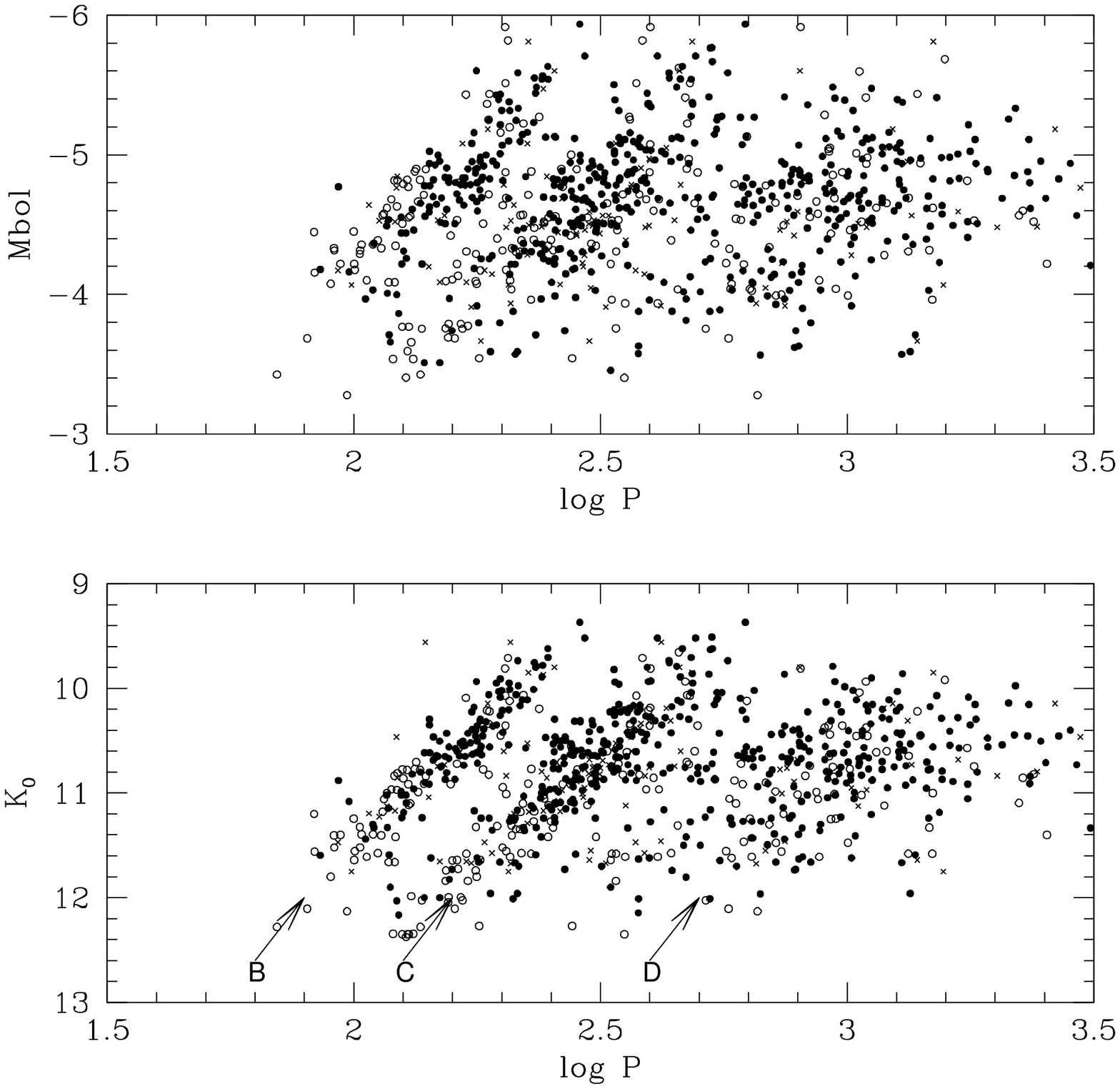}{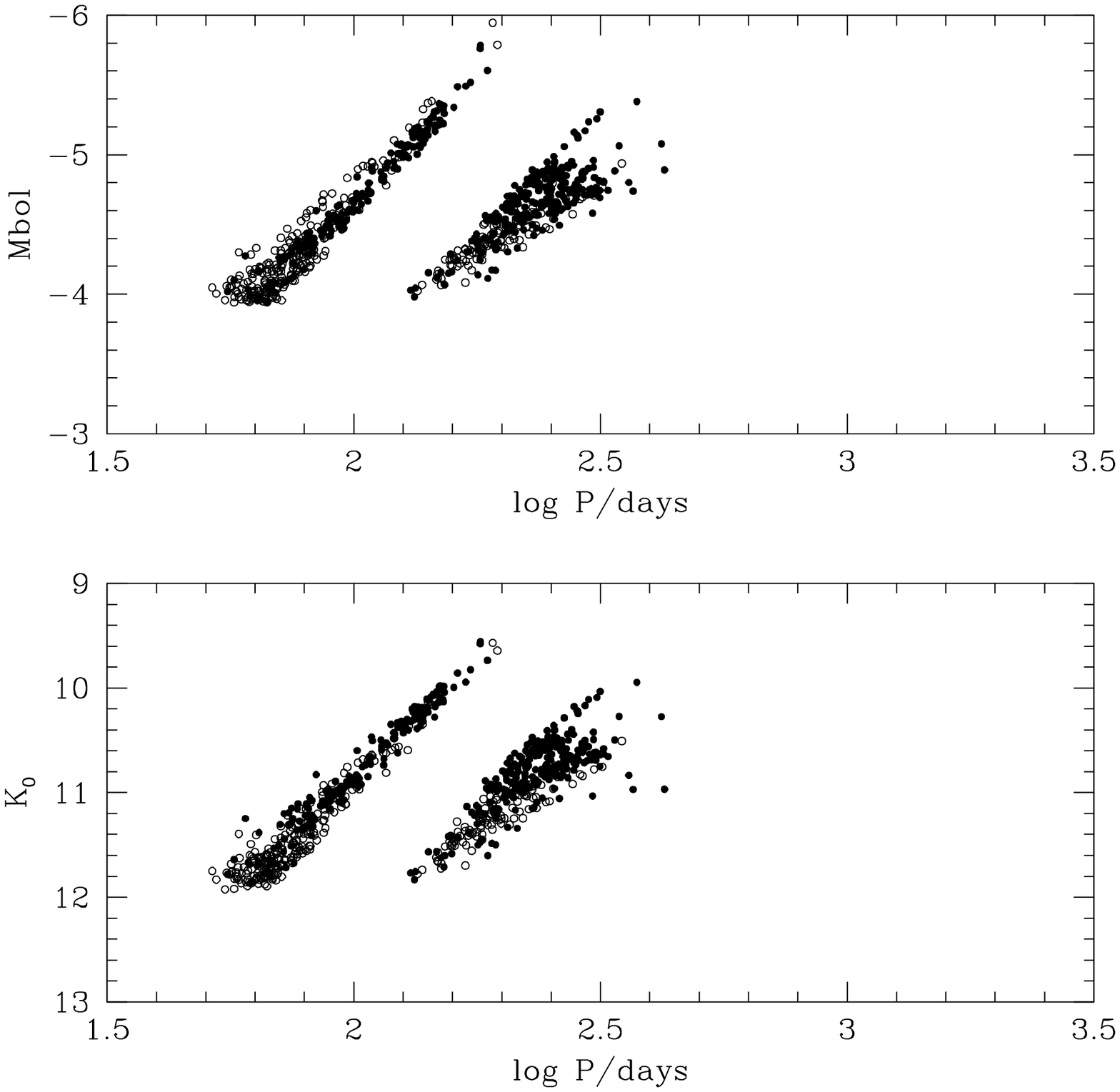}
\caption{Observed (left, data from Groenewegen 2004) and 
predicted (right) distribution of LMC TP-AGB stars in
period--luminosity diagrams (either in the K-band, or in $M_{\rm
bol}$). The observed sample has been limited in amplitude so that just
the sequences B, C and D are well delineated. The simulation includes
stars in both first overtone and fundamental mode, that correspond to
the observed sequences B and C, respectively.}
\label{fig_periods}
\end{figure}

Marigo et al.~(2003) already used some of these adaptations of
TRILEGAL to demonstrate that TP-AGB models with variable molecular
opacities do naturally produce the ``red tail'' of C-stars observed in
$K$ vs. $J\!-\!K$ diagrams of the MCs. Moreover, Cioni et al.~(2006ab,
and this meeting), used it to map variations of mean metallicity and
age across the MCs.  Figure~\ref{fig_mloss} and \ref{fig_periods}
present additional simulations for the stellar populations in the MCs
and in the Milky Way disk, that mostly regard the mass loss rates and
pulsation periods.

Figure~\ref{fig_mloss} shows the predicted distribution of mass-loss
rates for galactic M- and C-type Miras, as compared to
observations. It is interesting to note that the dispersion of
$\dot{M}$ at given period in this diagram is a consequence of using
Bowen \& Willson's (1991) and Wachter et al.'s (2002) prescriptions
for the mass loss (for M- and C-type stars, respectively), and it
would be absent if we were using simple fits to the empirical data,
like the relation provided by Vassiliadis
\& Wood (1993). On the other hand, tracks computed with Reimers' (1975)
mass loss formula would fail to reproduce the observed values at the
high end.
With this kind of simulations, one can really test whether present-day
prescriptions for mass loss (either theoretical or empirical) are
consistent with the observed distributions, taking into account also
the completeness of observed samples.

Figure~\ref{fig_periods} compares observed and simulated P--L
sequences of LPVs. So far, we have included just the first overtone
and fundamental modes of LPVs, that correspond to the B and C
sequences observed in microlensing surveys of the Magellanic
Clouds. We plot only the stars falling inside the LPV instability
strips (Gautschy 1999). Comparing with the data compiled by
Groenewegen (2004), we are able to reproduce some of the observed
features, such as the slopes of the sequences, the trend of C stars
being more luminous than M ones in sequences B and C, and the higher
fraction of C stars in sequence C (fundamental mode) than in B (first
overtone). However, first overtone periods are smaller than observed
in sequence B, and the detailed number counts in both sequences
suggest us that a small upward shift is needed to the ``critical
luminosities'' that define the transition from first overtone to
fundamental mode. Adjustments made to fix these problems would have a
moderate impact on the mass loss rates used in the models, that on the
other hand would affect properties like lifetimes and star counts. It
becomes then clear that more detailed comparisons of this kind are
worth being pursued.

In summary, we are ready to make statistically significant comparisons
between predictions of TP-AGB models, and observations of AGB star
samples in Local Group galaxies. These comparisons are no longer
limited to just the photometry, but now include other crucial
properties such as the mass loss rates, surface chemical composition,
and pulsation periods.


\acknowledgements 
This work was partially supported by the Padova University
(progetto di Ricerca di Ateneo CPDA052212).


\end{document}